\documentclass[prl,aps,twocolumn,showpacs,floatfix]{revtex4}
\usepackage{amsmath,graphics,epsfig,color,verbatim,ulem}

\begin{document}

\title{Epitaxial-strain-induced multiferroicity in SrMnO$_{3}$ from 
first principles}

\author{Jun Hee Lee}
\email{jhlee@physics.rutgers.edu}
\author{Karin M. Rabe}
\affiliation{Department of Physics and Astronomy,
Rutgers University, Piscataway, New Jersey 08854-8019, USA}

\begin{abstract}

First-principles density-functional calculations reveal a large 
spin-phonon coupling in cubic SrMnO$_{3}$, with ferromagnetic ordering 
producing a polar instability.
Through combination of this coupling with the strain-polarization 
coupling characteristic of perovskites, the bulk antiferromagnetic 
paraelectric ground state of SrMnO$_3$ is shown to be driven to a 
previously unreported multiferroic ferroelectric-ferromagnetic state by 
increasing epitaxial strain, both tensile and compressive.
This state has a computed polarization and estimated Curie temperature 
above 54 $\mu$C/cm$^2$ and 92 K. 
Large mixed magnetic-electric-elastic responses are predicted in the 
vicinity of the phase boundaries.

\end{abstract}

\maketitle

Multiferroic materials have been the subject of continuing attention 
both for fundamental physics and for potential applications including 
transducers and information
storage \cite{Spaldin05,Ramesh07,Cheong07}.
There is particular interest in the search for multiferroic materials 
with large polarization ($P>1\mu$C/cm$^2$) and magnetization that 
persists to high temperatures,
as well as a strong coupling between the magnetism and the electric 
polarization. This coupling may be enhanced by proximity of the system 
to a morphotropic phase boundary.

With epitaxial strain, it is possible to widen the search to include 
phases and phase boundaries that do not appear in bulk.
In many paraelectric perovskite oxides, epitaxial strain couples 
strongly to the lowest-frequency polar phonon; this is responsible for 
the phenomenon of epitaxial-strain-induced ferroelectricity, which has 
been intensively studied both experimentally and theoretically \cite{Haeni04,Antons05,Carl09}.
In an antiferromagnetic-paraelectric system, there is a further 
intriguing possibility. If the system has a spin-phonon coupling in 
which the lowest-frequency polar phonon is softer for ferromagnetic 
ordering than for antiferromagnetic ordering, then epitaxial strain 
enhancement of a polar instability lead to the lowering of the 
energy of the ferromagnetic (FM)-ferroelectric (FE) state below that of the 
antiferromagnetic (AFM)-paraelectric (PE) state. This mechanism for producing a 
multiferroic phase was proposed and elucidated using first-principles 
calculations for EuTiO$_3$ \cite{Rabe06}, with polarization above 30 $\mu$C/cm$^2$) 
at -2.0 \% strain but a low Curie temperature 
comparable to the bulk Neel temperature of 5.5 K \cite{ET}.

The demonstration of the spin-phonon coupling mechanism for 
epitaxial-strain-induced multiferroicity in EuTiO$_3$ suggests a search 
for other paraelectric antiferromagnetic materials in which this 
mechanism could be realized and the performance optimized.
The primary criterion is that of a large downward shift in the frequency 
of the lowest polar phonon with ferromagnetic ordering, which leads to a 
large energy gain for polar distortion of the ferromagnetic state.
Next, we look for systems with a moderate Neel 
temperature ($T_{\rm N}$): while it would be preferable 
to have the magnetic ordering temperatures as high as possible, this is 
limited by the fact that the FM-AFM energy splitting cannot be larger 
than the scale of the energy gain for polar distortion of the 
ferromagnetic state. Finally, the polar instability should be strong 
enough to compete with any other structural distortions, such as oxygen 
octahedron rotations.

$B$-site magnetic perovskite oxides are of particular interest, as the 
larger exchange coupling results in much higher magnetic ordering 
temperatures than that of $A$-site rare-earth systems such as EuTiO$_3$.
Perovskite manganites are especially promising as they show strong 
magnetoelectronic and magnetostructural effects, including colossal 
magnetoresistance \cite{CMR} and magnetic-field-induced structural 
transitions \cite{structural}. Indeed, a first-principles survey of the 
phonon dispersions of cubic perovskite oxides, including previously 
reported results on chromites \cite{Umesh08} and our own 
investigations of chromites, ferrites and manganites compounds \cite{ours} shows strong 
spin-phonon coupling in both SrMnO$_3$ and CaMnO$_3$. Though both compounds 
have moderate a moderate $T_{\rm N}$, we focus on SrMnO$_3$ as the strong oxygen octahedron 
instabilities in CaMnO$_3$ are less favorable for realization of the 
spin-phonon mechanism.

In this paper, we report first-principles observation of strong 
spin-phonon coupling and a resulting epitaxial strain induced 
multiferroic phase in SrMnO$_3$ (SMO), a AFM-PE $d^{\rm 3}$ perovskite oxide.  
In bulk, SMO is observed to have $G$-type antiferromagnetic ordering at 
$T_{\rm N}$=$\sim$233-260 K in the cubic perovskite structure \cite{smo,smo3}.
In the $G$-AFM phase, we find the epitaxial-strain-induced 
ferroelectricity previously suggested in Ref.~\onlinecite{Ghosez09}.
Further, we find that the polar instability is strongly coupled to 
ferromagnetic ordering. Thus, the increase in polarization produced by 
increasing tensile strain leads to
additional magnetic transitions, first to $C$-AFM and then to $A$-AFM 
ordering, with an increasing fraction of parallel nearest neighbor Mn 
spins, before the final transition to the ferroelectric FM phase. For 
compressive strain, the transition occurs directly from the $G$-AFM to the 
ferroelectric FM phase. While the cubic FM phase is metallic, in both 
cases, the system with the equilibrium polar distortion is insulating.
Finally, we predict large mixed magnetic-electric-elastic responses in 
the vicinity of the phase boundaries.

First-principles calculations were performed using density-functional 
theory within the
generalized gradient approximation GGA+$U$ method
with the Perdew-Becke-Erzenhof parametrization \cite{PBE}
as implemented in
the $Vienna$ $Ab$ $Initio$ $Simulation$ $Package$ 
(VASP-4.6)~\cite{Kresse2,Kresse3}; selected LSDA+$U$ calculations were 
performed for comparison.
We use the Dudarev~\cite{Dudarev} implementation
with on-site Coulomb interaction $U$=2.7 eV
and on-site exchange interaction $J_H$=1 eV
to treat the localized $d$ electron states in Mn.
Within GGA+$U$, this choice gives agreement between the calculated 
(2.7$\mu_B$) and experimental magnetic moments (2.6$\mu_B$$\pm$0.2) 
\cite{smo3} and is similar to
that in a previous GGA+$U$ study ~\cite{U1}.
The projector augmented wave (PAW) potentials \cite{Kresse1} explicitly
include 10 valence electrons for Sr (3$s2$3$p6$4$s2$), 13 for Mn
(3$p6$3$d5$4$s2$), and 6 for oxygen (2$s2$2$p4$).

The phonon frequencies of the ideal cubic perovskite $G$-AFM and FM 
reference structures were computed using the frozen phonon method in a 
$\sqrt 2\times \sqrt
2\times
2$ supercell
with a $6\times6\times4$ Monkhorst-Pack (M-P) $k$-point mesh
at the $\Gamma$, R, X and M points of the primitive
perovskite Brillouin zone; as spin-orbit coupling was not included, 
these wavevectors
remained good quantum numbers for both FM and $G$-AFM orderings.

\begin{table}
\caption{Calculated lowest phonon frequencies, in cm$^{-1}$, of cubic 
SrMnO$_3$
at calculated equilibrium lattice constants with $G$-AFM and FM orderings
for high symmetry $q$-points.}
\begin{ruledtabular}
\begin{tabular}{ccccc}
& $\Gamma$ & X & R & M \\
\hline
$G$-AFM ($a_0$=3.845\AA) & 121 & 116 & 84.5{\it i} & 38.1{\it i} \\
FM ($a_0$=3.845\AA) & 76.2{\it i} & 116 & 114{\it i} & 86.3{\it i} \\
FM ($a_0$=3.865\AA) & 109{\it i} & 113 & 119{\it i} & 89.9{\it i} \\
\end{tabular}
\end{ruledtabular}
\label{dispersion}
\end{table}
To find the minimum-energy configuration in a given space group 
determined by freezing in one or more unstable modes of the cubic 
reference structure, we moved the
atoms according to the conjugate-gradient algorithm until the residual 
Hellman-Feynman
forces were
less than 1.0meV/\AA. Structural optimizations were performed for 
20-atom $\sqrt 2\times \sqrt 2\times 2$ with a $4\times4\times4$ M-P 
$k$-point mesh;
for $A$-AFM R$_4^+$[110]+$\Gamma_4^-[110]$, a 2$\times$2$\times$2 
supercell with a $4\times4\times4$ M-P $k$-point mesh was used.

To study the effects of epitaxial strain, we performed ``strained-bulk'' 
calculations \cite{Pertsev} in which calculations were performed for the 
periodic crystal
with
appropriate epitaxial constraints imposed on the in-plane lattice 
parameters, with all atomic positions and the out-of-plane lattice 
constant optimized. Epitaxial
strain is here defined relative to the computed lattice constant for the 
$G$-AFM cubic perovskite structure (3.845\AA).
Ferroelectric polarizations for the relaxed structures at each strain
were computed by the Berry-phase method \cite{pol}.
Curie and Neel temperatures for a given strain are estimated from the 
energy differences between FM and $G$, $C$ and $A$-AFM orderings 
assuming two exchange constants
(in-plane and out-of-plane nearest neighbor couplings) and applying mean 
field theory \cite{mean-field}.
As exchange couplings are generally overestimated in DFT, to obtain a 
prediction useful for quantitative comparison with experiment, we then
uniformly rescaled the temperature so that the energy difference for the 
bulk cubic perovskite phase ($\Delta E$=86meV, $T_{N,MFT}$=3300K ) 
corresponds to the
experimental value of $T_N$=250K \cite{smo,smo3}.

The calculated lattice constants of cubic $G$-AFM and FM SrMnO$_3$ are
$a$=3.845\AA~and 3.865\AA, respectively; the slight overestimate 
relative to the experimental $G$-AFM value $a$=3.80\AA~\cite{smo,smo3} 
is typical of GGA calculations for oxides.
The computed lowest phonon frequencies of cubic FM and $G$-AFM SrMnO$_3$ 
at the computed lattice constant for the cubic $G$-AFM structure are 
shown in
Tab.~\ref{dispersion}; for the cubic FM structure, the computed lattice 
constant is very similar and the effect of the difference on the phonon 
frequency, also is
given in Table \ref{dispersion},
is small except at $\Gamma$.
The unstable modes at the M and R points are the oxygen octahedron 
rotations.
Across the Brillouin zone, the FM modes are lower in frequency than the 
corresponding modes in the $G$-AFM structure.
This effect is especially dramatic at $\Gamma$, where the lowest 
frequency TO mode is stable in the $G$-AFM structure but unstable in the FM 
structure.
The corresponding eigenvectors are (Sr,Mn,O$_{\parallel}$,O$_{\perp}$) = 
(0.10,0.31,-0.52,-0.56) for $G$-AFM and (0.03,0.41,-0.42,-0.57) for FM, 
showing displacement of
both
the Sr and Mn cations relative to a fairly rigid oxygen octahedral network.

 From the presence of unstable modes, it is clear that the computed ground
state of SrMnO$_3$ is not the ideal cubic perovskite structure.
However, the energy gains and distortions for the $G$-AFM structure 
resulting from freezing in the dominant M$_3^+$ and/or R$_4^+$ unstable 
modes are quite small. For
example,
the equilibrium R$_4^+$[001] rotation angle is 4.8$^\circ$, resulting in 
a $I$4/$mcm$ structure with an energy 6.2 meV/f.u. lower than the ideal 
cubic perovskite
structure.
$R$$\overline 3$$c$(R$_4^+$[111]) and $Imma$(R$_4^+$[110]) yield similar 
equilibrium angles and energy gains, while $P4/mbm$(M$_3^+$[001]) is 
smaller still, with
2.4$^\circ$ and energy gain 0.4 meV/f.u.
This is consistent with the experimental observation that AFM SrMnO$_3$ 
has a cubic structure but transforms to the $I$4/$mcm$ structure with a 
small fraction of
$A$-site substitution by the smaller cation Ca \cite{smo}.
For FM ordering, the energy of the equilibrium cubic structure is 76 
meV/f.u. higher than that of the $G$-AFM cubic structure,
and all FM structures are higher in energy than the lowest-energy AFM 
structure, despite the lower frequencies of the M, R and $\Gamma$ modes and
larger relaxation energies.

\begin{figure}
\begin{center}
\includegraphics[width=8.5cm,trim=0mm 8mm 0mm 0mm]{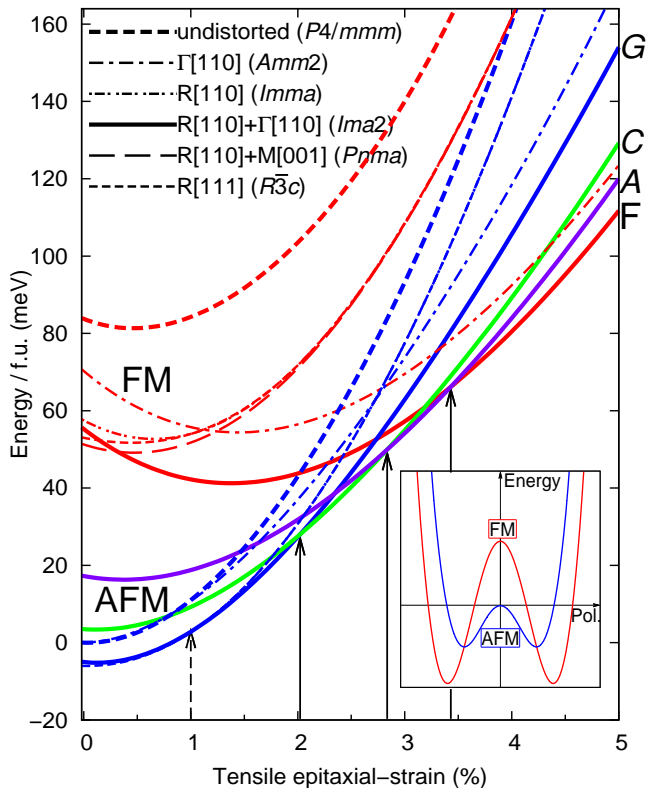}
\end{center}
\caption{(Color online) GGA+$U$
total energies of various structures, per formula unit, obtained by 
freezing in the given mode(s).
Calculations were performed at integer values of strain and interpolated.
The energies of structures with FM ordering are shown in red, for 
$G$-AFM in blue, $C$-AFM in green and $A$-AFM in violet.
Vertical black dotted lines at 1.0\%, 2.0\%, 2.8\% and 3.4\% strain 
indicate phase boundaries separating $Imma$($G$-AFM), $Ima$2($G$-AFM),
$Ima$2($C$-AFM), $Ima$2($A$-AFM) and $Ima$2(FM) states.
The inset is a schematic showing the stability of the FM-FE state at 
large strain.}
\label{energy}
\end{figure}

The restrictions on lattice vectors at 0\% epitaxial strain change these 
distortions and energy differences only slightly;
the results for the latter are included in Figure \ref{energy}.
For example, the tetragonal relaxation of the FM state lowers its energy 
slightly, so that the AFM-FM splitting is 84 meV/f.u. For the 
R$_4^+$[001] rotation, the
equilibrium angle is almost the same (4.7$^\circ$) and the energy 
lowering is slightly smaller (5.4 meV/f.u.).

In Fig.~\ref{energy}, the epitaxial strain dependence of the total 
energies of various structures and magnetic orderings is shown for 
tensile strain from 0\% to 5\%. The energies of the $G$-AFM phases are 
seen to increase with increasing tensile strain.
Because of their coupling to strain, the lowest
frequency polar
modes in the $G$-AFM $P$4/$mmm$ and $Imma$(R$_4^+$[110]) structures 
become unstable at a critical strain of about 1\%, with second-order 
phase transitions to the
polar
$Amm$2($\Gamma_4^-$[110]) and $Ima$2(R$_4^+$[110]+$\Gamma_4^-[110]$) 
phases, respectively. The polar distortion and associated energy gain 
increase with increasing
strain. This epitaxial-strain-induced ferroelectricity in the 
antiferromagnetic state
is analogous to that previously found in nonmagnetic $P$4/$mmm$ 
SrTiO$_3$\cite{Haeni04,Antons05}, $Pnma$ CaTiO$_3$ \cite{Carl09} and 
antiferromagnetic $Pnma$ CaMnO$_3$ \cite{Ghosez09}, with
critical strains of 0.6\%, 2\%, and 2\% respectively; hypothetical 
antiferromagnetic cubic BaMnO$_3$ \cite{Rondi09} is already 
ferroelectric at its equilibrium lattice constant.
The dependence of the total energies on compressive strain, not shown, 
produces an analogous effect at -2.9\% strain

Here, we focus instead on the interplay of the strain and polar 
instability with the magnetic ordering. In contrast to the AFM phases, 
the energy of FM structures initially decreases with increasing tensile 
strains (Fig.~\ref{energy}); this corresponds to the slightly larger 
computed lattice constant for FM ordering in the undistorted cubic 
structure. The polar instability leads to a substantial energy lowering 
which increases with increasing strain, so
that at tensile strains above 3.4\% the energy of the $Ima$2 FE-FM 
structure drops below those of the AFM-FE phases (see inset); for 
compressive strains, the FE-FM structure with space 
group $I$4$cm$(R$_4^+$[001]+$\Gamma_4^-[001]$) is favored 
above a critical strain of -2.9\%.
The polar
distortion with strain also drives a metal-insulator transition above a 
critical amplitude; at all strains considered, the FE-FM phase is 
insulating, with a band gap
ranging from 0.22 eV at 0\% strain to 0.51 eV at 5 \% strain.

In addition to $G$-AFM ordering, we also considered 
R$_4^+$[110]+$\Gamma_4^-[110]$ phases with $C$-AFM and $A$-AFM 
orderings, in which some fraction of nearest
neighbor Mn moments are
parallel. Though higher in energy than $G$-AFM at 0\% strain, they are 
favored by increasing tensile strain. As can be seen in Fig. 
\ref{energy}, the polar $C$-AFM
phase,
with 1/3 parallel-spin bonds, drops below $G$-AFM at 2 \%,
and the $A$-AFM phase drops with 2/3 parallel-spin bonds below the 
$C$-AFM phase at 2.8 \%.
For compressive strain, however, the distorted $G$-AFM state is lower in 
energy than the $A$ and $C$ states up to the strain where the FM state 
becomes lower in energy, and thus there are no intermediate transitions.

\begin{figure}
\begin{center}
\includegraphics[width=8.6cm,trim=0mm 8mm 0mm 0mm]{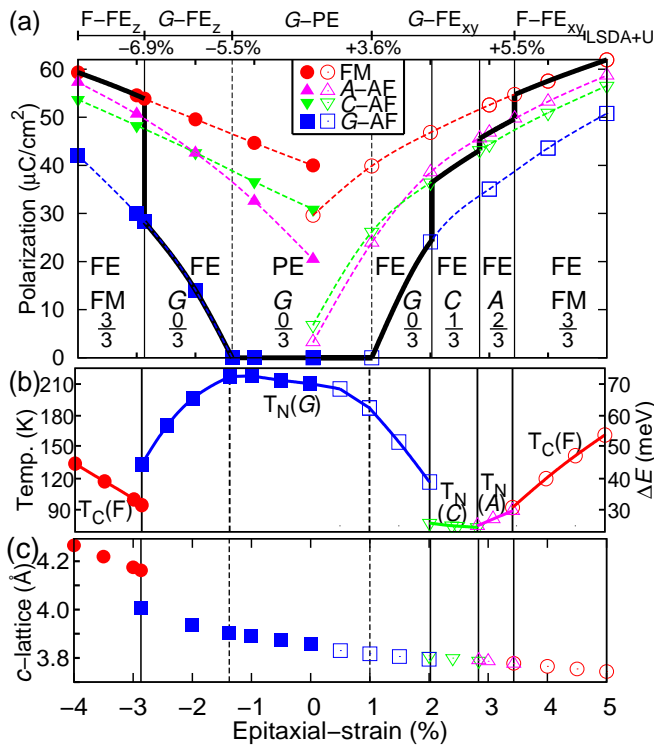}
\end{center}
\caption{(Color online) (a) Computed ferroelectric polarization and 
magnetic transition
temperature of SrMnO$_3$ $G$-AFM (square), $C$-AFM (down triangle), 
$A$-AFM (up triangle) and
FM (circle) in the lowest energy structure at each strain value.
The fractions of parallel nearest neighbor spins
in each state are given.
The bold black line follows the polarization
of the lowest energy structure with changing strain. The linear phase 
diagram at the top, not on the same scale, shows the phases and phase 
boundaries as computed
within LSDA+$U$ for comparison. Open and solid symbols represent structures where 
R rotation is along [110] and [001] respectively. 
(b) Magnetic ordering temperatures were estimated from computed values 
of $\Delta E$ as described in the text.
(c) $c$ lattice parameter for the lowest energy structure at each strain value.} 
\label{polarization}
\end{figure}
There are thus four phase transitions in the range of tensile strain 
considered: $G$-AFM-PE $\rightarrow$ $G$-AFM-FE $\rightarrow$ $C$-AFM-FE 
$\rightarrow$ $A$-AFM-FE
$\rightarrow$ FM-FE.
The strain dependence of the electric polarization and of the estimated 
Neel (for AFM phases) and
Curie (for the FM phase) temperatures are shown in Fig.~\ref{polarization}.
The three first-order magnetic transitions are of particular interest. 
At each of these, the change in magnetic order is accompanied by a 
change in the magnitude of
the
electric polarization. The magnitude of the
polarization change makes typical electric
energies ($P\cdot E$=10$\mu$C/cm$^2$ $\cdot$ 50 kV/cm=0.2 meV/f.u.) 
comparable to typical magnetic energies ($M\cdot H$=3$\mu$$_B$ $\cdot$ 
1T=0.2~meV/f.u.).
Thus, near the $A$-AFM-FE$\rightarrow$FE-FM phase boundary, an applied 
electric field can induce a nonzero magnetization. A
substantial magnetodielectric coupling is also expected at that phase 
boundary as a magnetic field induced transition to the FM phase will 
also lead to a jump in
the FE polarization and in the dielectric constant. At the AFM-FE/FM-FE 
phase boundary for compressive strain, in addition to an analogous 
magnetodielectric coupling, a strong strain response is expected due to 
the jump in $c$ lattice parameter.


As discussed above, this mechanism for epitaxial stabilization of a 
FM-FE phase based on spin-phonon coupling was first discussed for 
EuTiO$_3$ \cite{Rabe06}. In that
system,
the polar phonons for the bulk cubic phases are stable, with the phonon 
in the higher-energy FM phase about 10 cm$^{-1}$ lower in frequency than 
that in the $G$-AFM
phase. In contrast, in SrMnO$_3$ at 0\% strain both the FM-AFM energy 
difference and the FM-AFM difference in polar phonon frequency are much 
greater, the FM polar
phonon being unstable. As a result, the metastable FM-FE structure is 
already present at 0\% strain, though the electric field needed to 
induce this state
is inaccessibly high.
As in EuTiO$_3$, strain acts both to increase the ferroelectric 
instability in both phases and to lower the FM-AFM energy splitting. 
This is apparent in the lowering of $T_{\rm N}$ for $G$-AFM as the phase boundaries are approached, 
and in the increase of $T_{\rm C}$ with increasing magnitude of strain. 
The scale of the magnetic ordering temperatures at the phase
boundaries is set by the coupling of the magnetic energy splitting to the discontinuous
change in structure as the boundary is crossed; this coupling is so strong
that the magnetic ordering temperatures remain above 92 K throughout 
the strain range considered, even at
boundaries where the structural discontinuity is relatively
small. 
This offers operating temperatures higher than that of EuTiO$_3$ 
\cite{ET} and magnetically-driven ferroelectrics \cite{Kimura03,Andrei08}.

In GGA+$U$, the overestimate of the cell volume may lead to a spurious 
enhancement of the polar instability. We have investigated this effect 
by performing calculations
for the four lowest-energy phases with LSDA+$U$, where the underestimate 
of the volume is expected to lead to a comparable suppression of the 
polar instability. As can
be seen from the linear phase diagram at the top of Fig. 
\ref{polarization}, epitaxial-strain-induced ferroelectricity in the 
$G$-AFM phase moves to larger strain, and
the relative stability of the $C$-AFM and $A$-AFM phases decreases, so 
that the transition from $G$-AFM-FE to FM-FE occurs at a higher critical 
strain of 5.5\%, which can
be regarded as an upper bound. We thus estimate the critical s for the 
FM-FE phase to be +4.5 $\pm$ 1\% (tensile strain), and -4.9 $\pm$ 2\% 
(compressive strain).

It is well known that changes in the choice of the parameter $U$ can 
have a significant effect on magnetic ordering energies; this was shown 
in particular in
EuTiO$_3$ \cite{Rabe06}. We have verified that the 
epitaxial-strain-induced multiferroicity occurs for a wide range of 
possible choices of $U$, though the
quantitative details change. For example, use of a slightly larger value 
of $U$ reduces
the energy difference between FM and AFM phases while leaving the 
spin-phonon coupling relatively unchanged
so that the critical strain for the FM-FE phase shifts down from 3.4\%.  

The critical strain for observation of the FM-FE phase in 
SrMnO$_3$ is rather high, which may make experimental
confirmation challenging.
Observation of the behavior characteristic of lower strains, such as the 
strain-induced ferroelectricity in $G$-AFM and the decrease in T$_N$ in 
the $G$-AFM PE and FE
phases, should be taken as indicators of the impending transition to the 
FM-FE phase at higher strain.

In summary, we have presented first-principles density-functional 
calculations that reveal a large spin-phonon coupling in cubic 
SrMnO$_{3}$, with ferromagnetic
ordering producing a polar instability. Through combination of this 
coupling with the polarization-strain coupling characteristic of 
perovskite oxides, both tensile and compressive epitaxial strain drive 
the system through a series of phase transitions to a 
ferromagnetic-ferroelectric multiferroic state.
As the magnetic transitions are accompanied by a jump in electric 
polarization, there is the possibility of
electric field control of magnetic ordering; at the two boundaries 
between AFM and FM phases, the polarization can conversely be controlled 
by an applied magnetic field. At this boundary for compressive strain, 
the jump in $c$ lattice parameter also should yield a strong strain 
response to applied fields.
Though the cubic FM phase is metallic, the polar distortion opens a gap 
in the electronic density of states, resulting in the insulating 
character of the FM-FE phase.
This suggests that the search for epitaxial-strain-induced multiferroics 
could be productively extended to include other systems with metallic 
character for the FM
reference state.

We thank C.-J. Eklund, C. J. Fennie, V. Gopalan, D. R. Hamann, A. 
Malashevich, L. Palova, J. Rondinelli, D. Schlom, N. Spaldin, and D. 
Vanderbilt for valuable
discussions. This work was supported by MURI--ARO Grant
W911NF-07-1-0410 and ONR Grant N0014-00-1-0261. Part of this
work was carried out at the Aspen Center for Physics.


\end{document}